\documentclass[twocolumn,showpacs,preprintnumbers,amsmath,amssymb]{revtex4}
\usepackage{graphicx}
\usepackage{dcolumn}
\usepackage{bm}

\begin{document}

\title{Is $Y(4008)$ possibly a $1^{--}$ $\psi(3^3S_1)$ state?}
\author{Li-Jin Chen\footnote{reekingchen@live.cn}, Dan-Dan Ye and Ailin Zhang\footnote{Corresponding author:
zhangal@staff.shu.edu.cn}} \affiliation{Department of Physics,
Shanghai University, Shanghai 200444, China}


\begin{abstract}
The strong decays of the radially excited $\psi(3^3S_1)$ state are studied within the $^3P_0$ model. As a believed $\psi(3^3S_1)$, some strong decay widths and relevant ratios of $\psi(4040)$ are calculated in the model. The theoretical results are consistent with experiments. In a similar way, as a possible $\psi(3^3S_1)$, the same strong decay widths and relevant ratios of $Y(4008)$ are presented. Our study indicates that $Y(4008)$ is hard to be identified with a $\psi(3^3S_1)$ charmonium once it is confirmed under the $D^*\bar{D}^*$ threshold, but it is very possibly a $\psi(3^3S_1)$ charmonium once it is confirmed above the $D^*\bar{D}^*$ threshold by experiment.
\end{abstract}
\pacs{12.39.Jh;13.25.Gv;14.40.Pq\\
Keywords: $^3P_0$ model, strong decay}
\maketitle

\section{Introduction}
Since the discovery of $J/\Psi$, many charmonium and charmonium-like states have been observed~\cite{pdg12}. In these states, most of them are confirmed as $c\bar c$ charmonium states, some of them do not fit the predicted features of $c\bar c$ charmonium. Especially, in the past few years, some neutral ``X, Y" and charged``Z" resonances which cannot be simply accommodated in the $c\bar c$ picture have been observed and explored~\cite{pdg12}. How to understand and identify these resonances is a big challenge.

Several years ago, the Bell Collaboration observed a significant enhancement with mass $M=4008\pm40^{+114}_{-28}$ MeV and width $\Gamma=226\pm44\pm87$ MeV when measuring the cross section for $e^+e^-\to \pi^+\pi^-J/\psi$~\cite{belle2}. From its production, $Y(4008)$ has $J^{PC}=1^{--}$. There is a large uncertainty on the measured mass. $Y(4008)$ was not confirmed by the BaBar Collaboration~\cite{babar}.

Based on some analyses, the $\psi(3^3S_1)$ and $D^\star \bar D^\star$ molecular state possibility of $Y(4008)$ is studied in Ref.~\cite{Xiangliu08}. Through the calculated mass with the heavy quark-antiquark potential, $Y(4008)$ is suggested the $\psi(3^3S_1)$~\cite{BK}. In a one boson exchange model~\cite{ding}, the study does not support the interpretation of $Y(4008)$ as a $D^*\bar D^*$ molecule. In order to identify $Y(4008)$, it is interesting to study its strong decays in detail.

In fact, there is a $\psi(4040)$ which is commonly believed the $J^{PC}=1^{--}$ $\psi(3^3S_1)$~\cite{pdg12,TSE,swanson}. $\psi(4040)$ has mass and width
\begin{equation}
M=4039\pm1~\rm{MeV}, ~\Gamma=80\pm10 ~\rm{MeV}.
\end{equation}
The measured mass and total width of $\psi(4040)$ is consistent with theoretical predictions~\cite{TSE,swanson}.

Now the fact is that there are two states $\psi(4040)$ and $Y(4008)$, which are close to the threshold of $D^*\bar{D}^*$. Furthermore, these two states have different total decay widths. Even though the calculation of the strong decay of $\psi(4040)$ within the $^3P_0$ model has been performed in Ref.~\cite{TSE}, in order to find the difference and have a comparison, it will be interesting to study the strong decays of $\psi(4040)$ and $Y(4008)$ in the $^3P_0$ model at the same time.

The paper is organized as follows. After the introduction, the $^3P_0$ model is briefly reviewed and possible strong decay channels and decay amplitudes of the $\psi(3^3S_1)$ state are presented in Sec.II. In Sec. III, the numerical results in the $^3P_0$ model are obtained. The last section is devoted to a simple discussion and summary.

\section{$^3P_0$ model and possible charmonium strong decays of $\psi(4040)$ and $Y(4008)$}

Up to now, many strong decay models have been developed to describe the transition of hadrons to open-flavor final states. The $^3P_0$ model~\cite{micu1969,yaouanc1,yaouanc2} was first proposed by Micu~\cite{micu1969}, and further developed by Orsay Group~\cite{yaouanc1,yaouanc2}. In the model, the created quark-antiquark pair is supposed the vacuum quantum numbers $J^{PC}=0^{++}$. Although the intrinsic mechanism and the relation to the Quantum Chromodynamics are not very clear, the model is widely employed to study the OZI-allowed strong decays of a meson into two other mesons, as well as the two-body strong decays of baryons and other hadrons~\cite{capstick,PE,ackleh,TFPE,TNP,FE}.

\begin{figure}
\begin{center}
\includegraphics[height=4cm,angle=-180]{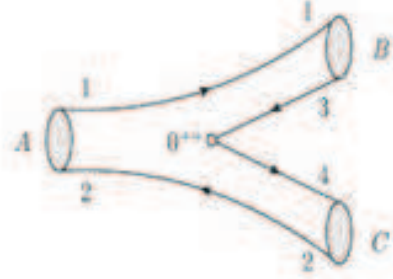}
\caption{The decay process of $A\Rightarrow B+C$ in the $^3P_0$ model~\cite{ZXX}.}
\end{center}
\end{figure}
A meson decay process $A\Rightarrow B+C$ is showed in Fig. 1. In the nonrelativistic limit, the transition operator is written as
\begin{eqnarray}\label{br1}
T=-3\gamma \sum_{m} \langle1m;1-m|00\rangle\int d\textbf{k}_{3}d\textbf{k}_{4}\delta^{3}( \textbf{k}_{3}+\textbf{k}_{4})\nonumber \\
\times y_{1m}(\frac{\textbf{k}_{3}-\textbf{k}_{4}}{2})\chi _{1, - m}^{34}\varphi^{34}_{0}\omega^{34}_{0} b^{\dagger}_{3i}( \textbf{k}_{3})d^{\dagger}_{4j}( \textbf{k}_{4})
\end{eqnarray}
where $i$ and $j$ denote the color indices for the $q\bar q$ pair. The flavor wave function for the $q\bar q$ pair is $\varphi^{34}_{0}\omega^{34}_{0}=(u\bar{u}+d\bar{d}+s\bar{s})/\sqrt{3}$, and $\omega^{34}_{0}=\delta_{ij}$ for the flavor and color singlet. $\chi _{1, - m}^{34}$ is the spin triplet. $y_{1m}(\textbf{k})=|\textbf{k}|\times Y_{1m}(\Omega)$ is the solid harmonic polynomial corresponding to the p-wave quark pair. The dimensionless constant $\gamma$ indicates the strength of the quark pair creation from the vacuum. Therefore, the helicity amplitude of the process $A\Rightarrow B+C$ reads as
\begin{widetext}
\begin{eqnarray}\label{maM}
 \mathcal{M}^{M_{J_A } M_{J_B } M_{J_C }} &=& \sqrt {8E_A E_B E_C } \gamma \sum_{M_{L_A } ,M_{S_A } ,M_{L_B } ,M_{S_B } ,M_{L_C } ,M_{S_C } ,m}\langle {1m;1 - m}|{00} \rangle \nonumber \\
 &&\times \langle {L_A M_{L_A } S_A M_{S_A } }| {J_A M_{J_A } }\rangle \langle L_B M_{L_B } S_B M_{S_B }|J_B M_{J_B } \rangle\langle L_C M_{L_C } S_C M_{S_C }|J_C M_{J_C }\rangle \nonumber \\
 && \times\langle\varphi _B^{13} \varphi _C^{24}|\varphi _A^{12}\varphi _0^{34} \rangle
\langle \chi _{S_B M_{S_B }}^{13} \chi _{S_C M_{S_C } }^{24}|\chi _{S_A M_{S_A } }^{12} \chi _{1 - m}^{34}\rangle I_{M_{L_B } ,M_{L_C } }^{M_{L_A } ,m} (\vec{K})
\end{eqnarray}
where $E_A=m_A$, $E_B =\sqrt {M_B^2 + \vec{K_B}^2 }$ and $E_C =\sqrt {M_B^2+ \vec{K_B}^2 }$ are the total energy of mesons $A$, $B$ and $C$. $\langle\varphi _B^{13} \varphi _C^{24}|\varphi _A^{12}\varphi _0^{34} \rangle $ and $\langle \chi _{S_B M_{S_B }}^{13} \chi _{S_C M_{S_C } }^{24}|\chi _{S_A M_{S_A } }^{12} \chi _{1 - m}^{34}\rangle$ are the matrix elements of favor wave functions and spin wave functions, respectively.
$I_{M_{L_B } ,M_{L_C } }^{M_{L_A } ,m} (\vec{K})$ is a spatial integral:
\begin{eqnarray}\label{I}
I_{M_{L_B } ,M_{L_C } }^{M_{L_A } ,m} (\vec{K}) &=& \int d \vec{k}_1 d \vec{k}_2 d \vec{k}_3 d \vec{k}_4 \delta ^3 (\vec{k}_1 + \vec{k}_2)\delta ^3 (\vec{k}_3+ \vec{k}_4)\delta ^3 (\vec{k}_B- \vec{k}_1- \vec{k}_3 )\nonumber \\
&&\times \delta ^3 (\vec{k}_C- \vec{k}_2 -\vec{k}_4) \Psi _{n_B L_B M_{L_B } }^* (\vec{k}_1 ,\vec{k}_3)\Psi _{n_cL_C  M_{L_c}}^* (\vec{k}_2 ,\vec{k}_4)\nonumber \\
&& \times \Psi _{n_A L_A M_{LA}} (k_1 ,k_2 )Y _{1m}\left(\frac{\vec{k_3}-\vec{k}_4}{2}\right).
\end{eqnarray}

Using the Jacob-Wick formula~\cite{JW}, the helicity amplitude can be transformed into the partial wave amplitude:
\begin{eqnarray}
\mathcal{M}^{JL} (A \to BC) &=& \frac{{\sqrt {2L + 1} }}{{2J_A  + 1}}\sum{M_{J_B } ,M_{J_C } } \langle {L0JM_{J_A } } |{J_A M_{J_A } }\rangle \nonumber \\
&&\times \left\langle {J_B M_{J_B } J_C M_{J_C } } \right|\left. {JM_{J_A } } \right\rangle M^{M_{J_A } M_{J_B } M_{J_C } } (\vec{K})
\end{eqnarray}
\end{widetext}
where $\vec{J}=\vec{J_B}+\vec{J_C}$, $\vec{J_A}=\vec{J_B}+\vec{J_C}+\vec{L}$, $M_{J_A}=M_{J_B}+M_{J_C}$.
The decay width is thus obtained as
\begin{eqnarray}
\Gamma  = \pi ^2 \frac{|\vec{K}|}{M_A^2}\sum{JL} |{\mathcal{M}^{JL}}|^2
\end{eqnarray}
where $|\vec{K}|$ is the momentum of the daughter meson in the initial meson A's center mass frame
\begin{eqnarray}
 |\vec{K}|= \frac{{\sqrt {[m_A^2  - (m_B  - m_C )^2 ][m_A^2  - (m_B  + m_C )^2 ]} }}{{2m_A }}.
\end{eqnarray}

With these formula in hand, we proceed with the study of the strong decays of $\psi(4040)$ and $Y(4008)$. $\psi(4040)$ and $Y(4008)$ have the same quantum number $J^{PC}=1^{--}$. Once they are assigned as the $1^{--}$ $\psi(3^3S_1)$ state, all possible open-charm strong decay modes allowed by the OZI rule above the $D^*\bar{D}^*$ threshold are given in Table. I. Accordingly, the decay amplitudes and the detailed decay channels are presented.

\begin{table}
\caption{The allowed open-charm strong decays of $1^{--}$ $\psi({3^{3}S_{1}})$ for $\psi(4040)$ and $Y(4008)$, where $\varepsilon=\gamma\sqrt{E_AE_BE_C}$.} \label{table-1}
\begin{tabular}{cccccccc}
\hline\hline
State & Decay mode & Decay amplitude & Decay channel\\
\hline
$                $    & $0^{-}+ 0^{-}$  & $\mathcal{M}^{00}=-\frac{\sqrt3}{18}\gamma\sqrt\varepsilon I_{00}$ & $D\bar{D},D_s\bar{D}_s$  \\
$\psi(3^{3}S_{1})$    & $0^{-}+ 1^{-}$  & $\mathcal{M}^{11}=-\frac{\sqrt6}{18}\gamma\sqrt\varepsilon I_{00}$ & $D\bar{D}^\star/\bar{D}D^\star$\\
$                $    & $1^{-}+1^{-}$   & $\mathcal{M}^{21}=-\frac{\sqrt5}{9}\gamma\sqrt\varepsilon I_{00}$ & $D^\star \bar{D}^\star$ \\
\hline\hline
\end{tabular}
\end{table}

For the flavor matrix element $\langle\varphi _B^{13} \varphi _C^{24}|\varphi _A^{12}\varphi _0^{34} \rangle$, there are several definitions which give different numbers. In our calculation, $\langle\varphi _B^{13} \varphi _C^{24}|\varphi _A^{12}\varphi _0^{34} \rangle=\frac{1}{\sqrt3}$ is chosen.

\section{Numerical results}

In order to get the numerical results within the $^3P_0$ model, several parameters are chosen as follows. The masses of constituent quarks are taken as $m_u=m_d=0.22~\rm{GeV}$, $m_s=0.419~\rm{GeV}$ and $m_c=1.6~\rm{GeV}$~\cite{XZZ}. The masses of relevant charmed mesons~\cite{pdg12} are listed in Table. II, where $(\pm)$ indicates the charged mesons and $(0)$ indicates the charge neutral mesons.
\begin{table*}
\begin{center}
\caption{The relevant mass and $R$ values of charmed mesons used in our calculation}
\begin{tabular}{cccccc}
\hline
\hline
Meson &$D $ &$D^\ast$ &$D_s$&$D_s^\ast$      \\
\hline
Mass (MeV) & 1869.62 ($\pm$), 1864.84 (0)    &2010.29 ($\pm$), 2006.97 (0)    &1968.49 ($\pm$)    &2112.3 ($\pm$) \\
R($\rm{GeV}^{-1}$)~\cite{YZJ,PRP}&1.52& 1.85& 1.41& 1.69\\
\hline\hline
\end{tabular}
\label{table21}
\end{center}
\end{table*}

There are other two important parameters in the $^3P_0$ model, the strength of quark pair creation $\gamma$ and the $R$ value in the simple harmonic oscillator (SHO) wave function. For the color saturation, the color matrix element as a constant can be absorbed into the dimensionless constant $\gamma$. In our calculation, $\gamma$ is chosen with $\gamma=6.3$~\cite{XZZ,YZJ,li,zhang} and the strength of $s\bar s$ pair creation $\gamma_{ss}=\gamma/\sqrt3$~\cite{yaouanc2}. The chosen $\gamma=6.3$ has a factor $\sqrt{96\pi}$ difference with that in Ref.~\cite{TSE}. The $R$ value in the SHO wave function can be obtained from the Schrodinger equation within the potential model~\cite{SR}.

In general, there are two ways to choose $R$: a constant around $2$ $\rm{GeV}^{-1}$~\cite{PE,TSE,chen} and an effective varying value ~\cite{FE,li}. In this paper, an effective $R$ is chosen and the suitable region of $R$ is fixed by $\psi(4040)$. At that $R$, the strong decay widths and relevant ratios of $Y(4008)$ are investigated. Of course, the numerical results depend on $R$. To learn this dependence, the variation of our results with $R$ are also presented.

\subsection{$\psi$(4040)}

\begin{figure}
\begin{center}
\includegraphics[height=4.3cm]{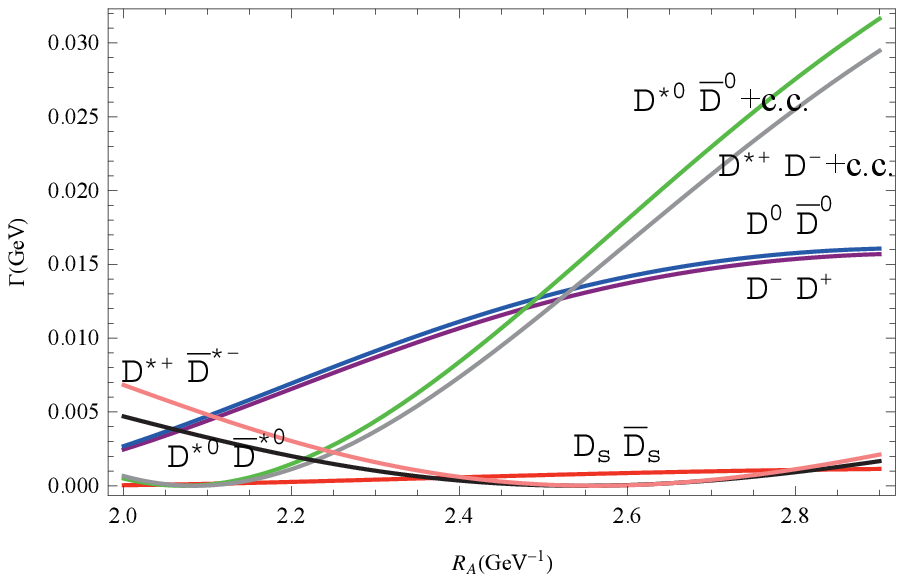}
\includegraphics[height=4.3cm]{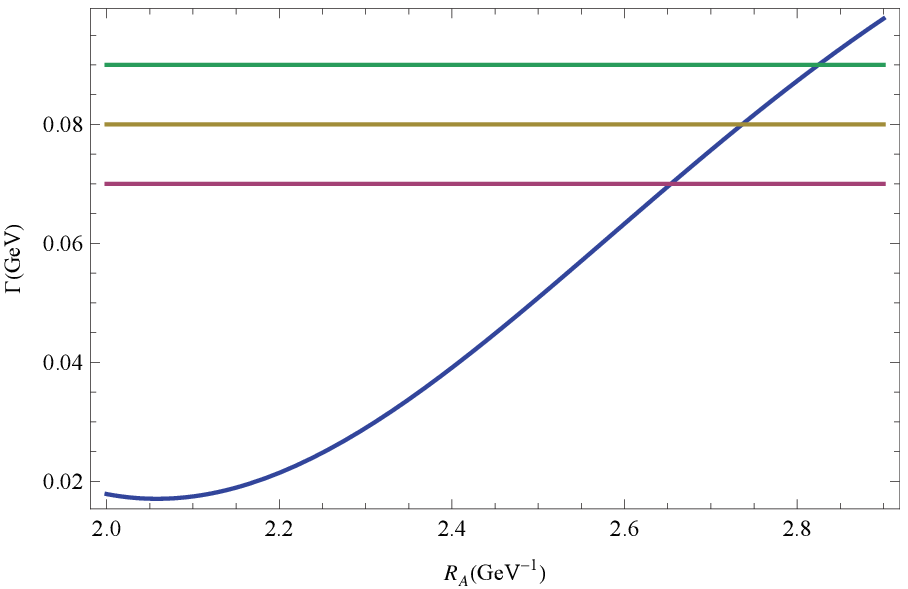}
\caption{(color online)(a) Possible partial strong decay widths of $\psi(4040)$ versus $R$; (b) The total strong decay width of $\psi(4040)$ versus $R$.}
\end{center}
\end{figure}

As a commonly believed $\psi(3^3S_1)$, the variation of the decay width of $\psi(4040)$ for different modes with $R_A$ is shown in Fig. 2(a). The variation of the total decay width of $\psi(4040)$ with $R_A$ is presented in Fig. 2(b). From PDG~\cite{pdg12}, three horizontal lines in the figure are drawn to indicate the lower, central and upper values of the total width of $\psi(4040)$ ($\Gamma=0.08\pm0.01$ GeV). $R_A$ (corresponding to the initial $A$ meson) is therefore fixed by the three lines at the region $2.65\to 2.82$ GeV$^{-1}$ with the central value $2.74$ GeV$^{-1}$. At $R_A=2.74$ GeV$^{-1}$, the widths of all possible open-flavor strong decay channels are calculated and given in Table. III. As a comparison, the results in Ref.~\cite{TSE} are also listed. Obviously, the dominant decays of $\psi(4040)$ are $D\bar{D}$, $D^*\bar{D}$ and $D^*\bar{D}^*$ channels.
\begin{table*}
\centering
\caption{Open-flavor strong decays of $\psi(4040)$ at universal $R_A=2.74$ GeV$^{-1}$ (in MeV)}
\label{table}
\begin{tabular*}{16cm}{@{\extracolsep{\fill}}ccccccc}
  \hline
  \hline
  Decay Channles & $D\overline{D}$ & $D_s\overline{D}_s$ & $D\overline{D}^*/\overline{D}D^*$ & $D^*\overline{D}^*$ & $\Gamma(total)_{thy}$ & $\Gamma(total)_{expt}$ \\
  \hline
  Ref.~\cite{TSE} &  0.1     & 7.8  & 33 & 33  & 74   & 80$\pm$10 \\
  Our results &  30.44     & 1.02  & 47.67 & 0.87  & fixed point   & 80$\pm$10 \\
  \hline
  \hline
\end{tabular*}
\end{table*}

Unlike the decay widths, the ratios of the decay widths are less sensitive to the uncertainties of the $^3P_0$ model. Therefore, some relevant ratios are calculated and presented in Table. IV. The experimental data are those from PDG~\cite{pdg12}.

Except for $\frac{\Gamma(D^*(2007)^0\bar{D}^*(2007)^0)}{\Gamma(D^*(2007)^0\bar{D}^0+c.c.)}$ and $\frac{\Gamma(D^0\bar{D}^0)}{\Gamma(D^*(2007)\bar{D}^0+c.c.)}$ (measure in 1977~\cite{gold}), our results are consistent with experiments. Besides, the $\mathcal{BR}(\psi(4040)\Rightarrow D\bar{D})$ is $37.5^{+3.9}_{-3.1}\%$, which is also consistent with the BABAR data $(31.2\pm5.3)\%$~\cite{BaBar} within the experimental uncertainty. In Ref.~\cite{HX}, the obtained $\mathcal{BR}(\psi(4040) \Rightarrow D\bar{D})=(25.3\pm4.5)\%$.
\begin{table*}[htbp]
  \centering
  \caption{Relevant ratios of $\psi(4040)$ at $R_A=2.74$ GeV$^{-1}$}
    \begin{tabular}{cccccc}
    \toprule
    Ratios & $\frac{\Gamma(D\bar{D})}{\Gamma(D^*\bar{D}+c.c.)}$     & $\frac{\Gamma(D^*\bar{D}^*)}{\Gamma(D^*\bar{D}+c.c.)}$     & $\frac{\Gamma(D^*(2007)^0\bar{D}^*(2007)^0)}{\Gamma(D^*(2007)^0\bar{D}^0+c.c.)}$     & $\frac{\Gamma(D^*(2010)^+D^-)}{\Gamma(D^*(2007)^0\bar{D}^0+c.c.)}$     & $\frac{\Gamma(D^0\bar{D}^0)}{\Gamma(D^*(2007)\bar{D}^0+c.c.)}$ \\
    \hline
    Expt. & $0.24\pm0.05\pm0.12 $& $0.18\pm0.14\pm0.03$ & $32\pm12$ & $0.95\pm0.09\pm0.10$ & $0.05\pm0.03$ \\
    our results & 0.641 & 0.018 & 0.022 & 0.927 & 0.625 \\
    Ref.~\cite{TSE} & 0.003 & 1     &       &       &   \\
    \hline
    \hline
    \end{tabular}
  \label{tab:addlabel}
\end{table*}

In our results, $\frac{\Gamma(D\bar{D})}{\Gamma(D^*\bar{D}+c.c.)}$ and $\frac{\Gamma(D^0\bar{D}^0)}{\Gamma(D^*(2007)\bar{D}^0+c.c.)}$ have pole around $R_A=2.07$ GeV$^{-1}$ as pointed out in Refs.~(\cite{SN,PE,RN}).

\subsection{$Y(4008)$}

As indicated in the first section, $Y(4008)$ is close to the $D^*\bar{D}^*$ threshold while has a large mass uncertainty. Therefore, more decay channels may open when $Y(4008)$ has a larger mass. To check the possibility of $\psi(3^3S_1)$, the $R_A=2.74$ GeV$^{-1}$ fixed by $\psi(4040)$ is employed to study the open-flavor strong decay of $Y(4008)$. At $R_A=2.74$ GeV$^{-1}$, the widths of all possible open-flavor decay channels are presented in Table. V, where $3940$, $4008$ and $4162$ represent the lower, central and upper mass of $Y(4008)$, respectively.

\begin{table*}
\centering
\caption{Open-flavor strong decays of $Y(4008)$ at $R_A=2.74$ GeV$^{-1}$ (in MeV)}
\label{table}
\begin{tabular*}{16cm}{@{\extracolsep{\fill}}ccccccc}
  \hline
  \hline
  Decay Channles & $D\overline{D}$ & $D_s\overline{D}_s$ & $D\overline{D}^*/\overline{D}D^*$ & $D^*\overline{D}^*$ & $\Gamma(total)_{thy}$ & $\Gamma(total)_{expt}$ \\
  \hline
  ~~~~~~~~~~~3940 MeV & 17.9  & 7.7$\times10^{-6}$  & 8.61   & -   & 26.53    & \\

  $Y(4008)$ 4008 MeV &  26.44  & 0.59 & 32.18  &  -        & 59.21   & $226\pm44\pm87$\\

  ~~~~~~~~~~~4162 MeV &  43.44  & 3.07 & 120.36  & 59.02 & 225.89 &   \\
  \hline
  \hline
\end{tabular*}
\end{table*}

From Table. V, it is easy to find that $Y(4008)$ with the lower or central mass does not open the $D^*\overline{D}^*$ channel. Therefore, the predicted total decay width of $Y(4008)$ is largely different from the observed one. If $Y(4008)$ has the upper mass, the $D^*\overline{D}^*$ channel opens and the predicted total decay width is consistent with experiment. To learn the dependence of the total width of $Y(4008)$ on $R_A$, two figures corresponding to the central and upper mass are drawn in Fig. 3, where the horizontal lines indicate the experimental result.

\begin{figure}
\begin{center}
\includegraphics[height=4.3cm]{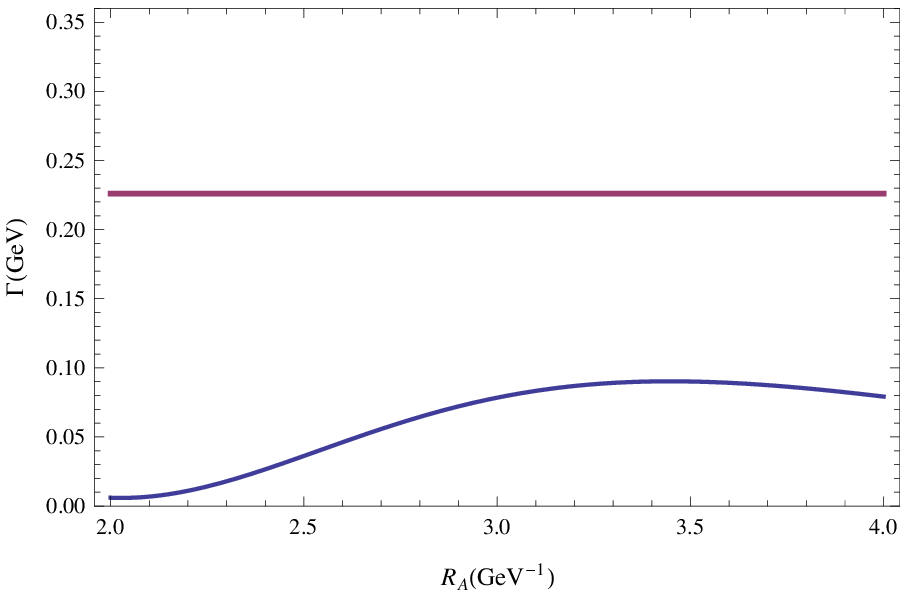}
\includegraphics[height=4.3cm]{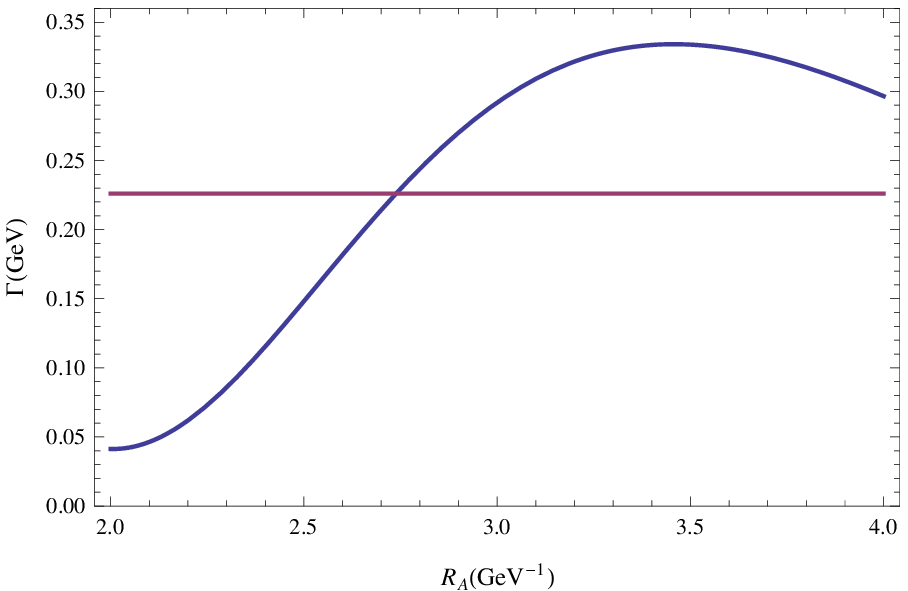}
\caption{(color online) Total decay widths of $Y(4008)$ versus $R_A$ for: (a) the central mass $4008$ MeV; (b) the upper mass $4162$ MeV.}
\end{center}
\end{figure}

Similarly, relevant ratios of $Y(4008)$ are calculated and presented in Table. VI. Unfortunately, there is no such experimental data at present.
\begin{table*}[htbp]
  \centering
  \caption{Relevant ratios of $Y(4008)$ with upper mass at $R_A=2.74$ GeV$^{-1}$}
    \begin{tabular}{cccccc}
    \toprule
    Ratios & $\frac{\Gamma(D\bar{D})}{\Gamma(D^*\bar{D}+c.c.)}$     & $\frac{\Gamma(D^*\bar{D}^*)}{\Gamma(D^*\bar{D}+c.c.)}$     & $\frac{\Gamma(D^*(2007)^0\bar{D}^*(2007)^0)}{\Gamma(D^*(2007)^0\bar{D}^0+c.c.)}$     & $\frac{\Gamma(D^*(2010)^+D^-)}{\Gamma(D^*(2007)^0\bar{D}^0+c.c.)}$     & $\frac{\Gamma(D^0\bar{D}^0)}{\Gamma(D^*(2007)\bar{D}^0+c.c.)}$ \\
    \hline
    our results & 0.361 & 0.492 & 0.505 & 0.964 & 0.357 \\
    \hline
    \hline
    \end{tabular}
  \label{tab:addlabel}
\end{table*}

\section{Summary and discussion}
In this work, the strong decay of the $1^{--}$ $\psi(3^3S_1)$ resonance is studied in the $^3P_0$ model. As a commonly believed $\psi(3^3S_1)$, the dominant strong decay of $\psi(4040)$ are $D\bar{D}$, $D^*\bar{D}$ and $D^*\bar{D}^*$ channels. Accordingly, the decay widths of these channels are calculated. Based on these decay widths, some relevant ratios are obtained. Most of the ratios are consistent with experiments within the experimental uncertainties. Our results for $\frac{\Gamma(D^*(2007)^0\bar{D}^*(2007)^0)}{\Gamma(D^*(2007)^0\bar{D}^0+c.c.)}$ and $\frac{\Gamma(D^0\bar{D}^0)}{\Gamma(D^*(2007)\bar{D}^0+c.c.)}$ are different from experimental data which were measured in 1977. Of course, the uncertainties related to the $^3P_0$ model are not studied in this paper, which may bring in some uncertainties.

$Y(4008)$ are close to the threshold of $D^*\bar{D}^*$ and has a large mass uncertainty. For this reason, the strong decays of $Y(4008)$ with different mass are studied. Under the threshold of $D^*\bar{D}^*$, it is hard to understand the wide decay width of $Y(4008)$ if $Y(4008)$ is assumed as the $\psi(3^3S_1)$. However, above the threshold of $D^*\bar{D}^*$, $Y(4008)$ is very possibly the $\psi(3^3S_1)$. In this case, more information is required to distinguished $\psi(4040)$ from $Y(4008)$ both in theory and in experiment.

To have a clear picture of the charmonium spectroscopy, the observed $X,~Y$ and $Z$ have to be understood and identified. Unfortunately, people has not a comprehensive understanding of these resonances. Besides, $Y(4008)$ was observed only by the Belle collaboration, and only the total decay width was given. More experiments are required to confirm its existence or not. Especially, the mass uncertainty of $Y(4008)$ has to be deduced if it is confirmed in forthcoming experiment. Only when more decay channels and their branching fractions ratios have been measured, can we understand $Y(4008)$ and $\psi(4040)$.

\begin{acknowledgments}
This work is supported by National Natural Science Foundation of China(11075102) and the Innovation Program of Shanghai Municipal Education Commission under grant No. 13ZZ066.
\end{acknowledgments}

\end{document}